\documentclass[a4paper]{article}

\usepackage{INTERSPEECH2019}
\usepackage{hyperref}
\usepackage{enumitem}
\usepackage{graphicx}
\usepackage{multirow}
\usepackage{multicol}

\newcommand{\z}{\mathbf{z}}
\newcommand{\uu}{\mathbf{u}}
\newcommand{\sigmab}{\boldsymbol{\sigma}}
\newcommand{\mub}{\boldsymbol{\mu}}

\newcommand{\kmeans}{\textsc{AE\textsubscript{K--MEANS}}}
\newcommand{\vamp}{\textsc{VAE\textsubscript{VAMP}}}

\makeatletter
\def\blfootnote{\xdef\@thefnmark{}\@footnotetext}
\makeatother

\title{Perception of prosodic variation for speech synthesis \\ using an unsupervised discrete representation of F0}

\name{Zack Hodari, Catherine Lai, Simon King}
\address{Centre for Speech Technology Research, University of Edinburgh, UK}
\email{\{zack.hodari, c.lai, Simon.King\}@ed.ac.uk}

\begin{document}

\maketitle

\begin{abstract}
In English, prosody adds a broad range of information to segment sequences, from information structure (e.g.\ contrast) to stylistic variation (e.g.\ expression of emotion). However, when learning to control prosody in text-to-speech voices, it is not clear what exactly the control is modifying. Existing research on discrete representation learning for prosody has demonstrated high naturalness, but no analysis has been performed on what these representations capture, or if they can generate meaningfully-distinct variants of an utterance. We present a phrase-level variational autoencoder with a multi-modal prior, using the mode centres as `\emph{intonation codes}'. Our evaluation establishes which \emph{intonation codes} are perceptually distinct, finding that the \emph{intonation codes} from our multi-modal latent model were significantly more distinct than a baseline using k-means clustering. We carry out a follow-up qualitative study to determine what information the \emph{codes} are carrying. Most commonly, listeners commented on the \emph{intonation codes} having a statement or question style. However, many other affect-related styles were also reported, including: emotional, uncertain, surprised, sarcastic, passive aggressive, and upset. Finally, we lay out several methodological issues for evaluating distinct prosodies.

\end{abstract}

\section{Introduction} \label{sec:introduction}

In text-to-speech synthesis (TTS), the natural variability of prosody is often not accounted for. Current TTS systems default to the production of average prosody \cite{zack-SSW19:2019}: monotonous and boring speech. Synthetic voices do not take contextual variation into account during training, thus different prosodies are seen as noise and only the mean is learnt. This sort of overly smoothed speech can be fatiguing to listen to in long form speech. However, relevant context can be very wide ranging and much of this can be expensive or impractical to obtain.  For example, previous work has identified consistent variations in prosody with respect to structural elements in the discourse context \cite{farrus_paragraph-based_2016,Cole2016,Kleinhans2017} and sentence level information structure \cite{calhoun_centrality_2010,lai2012response}. Variation has also been identified with respect to specific speaker attitudes \cite{armstrong2015contribution,lai2010uncertain,Betz2019} and stances \cite{Hubscher2018,freeman2019prosodic, ward2018inferring}.

Given this, a successful TTS system should be able to produce a large variety of plausible prosodic forms for a given utterance.  However, current TTS systems often rely on pre-specified linguistic context features to guide prosodic realisations of synthetic speech. Unsurprisingly, annotated speech data with wide coverage of suitably-rich contextual information is not widely available. So, in order to develop TTS models that generate plausible and appropriate prosody given a specific context, we propose to split the problem in half: \emph{controllability}---designing a system that can produce distinct renditions of isolated sentences; and \emph{appropriateness}---choosing the most appropriate rendition using contextual information. This paper presents work on the first task: learning a prosodic representation capable of producing distinct renditions of a single sentence. Importantly, we do not attempt to identify the most appropriate or most likely prosody given a pre-specified context. Instead, our goal is to verify that different renditions produced by our representation are \emph{perceived} as distinct, and whether they convey different information or intent.

Most TTS research on controllability focuses on emotion or emphasis \cite{yamagishi-emotion:2004,gustav-GMMQ-VAE:2018}. Conversely, more fundamental prosody research has focused on how acoustic-phonetic features map to linguistic categories. We want to bridge this gap in order to make advances in both together, by determining what meaning listeners perceive in renditions from controllable TTS. Recent phonetic studies support the idea that both categorical and continuous features are integral to prosodic variation  \cite{grice2017integrating,cole-RPT:2017}. In line with this, we learn \emph{discrete} representations which can potentially capture categorical differences often associated with phrasing and prominence, but also allow for the generation of fine-grained phonetic differences, which vary the perception of expressivity, emphasis, and speaker affect.

We use a variational autoencoder with a multi-modal prior (Section~\ref{sec:VAMP}) to learn a discrete representation of F0. We evaluate what these `\emph{intonation codes}' capture through subjective (Section~\ref{sec:evaluation-subjective}) and qualitative (Section~\ref{sec:evaluation-qualitative}) tests.

\section{Related work} \label{sec:related_work}

Controllable TTS has been approached from both supervised and unsupervised perspectives. Henter et al.\ \cite{gustav-GMMQ-VAE:2018} demonstrated that both can achieve the same quality for emotion control.

Unsupervised representation learning in TTS typically uses a continuous representation (i.e.\ $\mathbb{R}^n$) at the sentence level \cite{oliver-control:2015,CHiVE:2019,zack-SSW19:2019}, but this becomes increasingly difficult to interpret for $n \gtrsim 3$. Poor interpretability limits the range of practical use cases. For example, \cite{tacotron-GST:2018,CHiVE:2019} are limited to transferring style from another natural utterance. To address practical limitations, high-dimensional representations can be predicted automatically, perhaps using the current utterance \cite{tacotron-TP-GST:2018,amazon-dynamic-prosody:2020}. Discrete representations are another way to address interpretability \cite{srikanth-template:2016}, and can also be paired with automatic prediction from text \cite{xin-F0-VQ-VAE:2019}.

Prosody should be modelled in the correct \emph{domain}. While most approaches \cite{oliver-control:2015,tacotron-GST:2018,gustav-GMMQ-VAE:2018,CHiVE:2019,zack-SSW19:2019} operate on sentences, the sentence domain may not be the most appropriate for a fixed-sized prosodic representation. For example, sentences contain a variable number of prosodic phrases. It is likely that by working in the sentence domain, something closer to sentence-style, as opposed to prosody, is captured. Much less work has been done on prosodically-appropriate domains. Wang et al.\ \cite{xin-F0-VQ-VAE:2019} compare a discrete representation of F0 in the phrase domain to smaller and longer domains. Reconstruction performance clearly shows that these fixed-sized representations are less accurate for longer domains which can contain more information.

Although claims of expressivity or prosody control are often made, variability or controllability are often not evaluated. In \cite{CHiVE:2019}, prosody reconstruction measures the model's top-line performance, and prosody transfer is demonstrated qualitatively, but interpreting the latent space, or choosing the best rendition was not tackled. Tyagi et al.\ \cite{amazon-dynamic-prosody:2020} present a unit selection-like system for prosody generation. Prosody embeddings of the training sentences act as templates and are chosen using a linguistically-informed target cost and an acoustic join cost. They evaluate general appropriateness of isolated sentences using linguistic expert listeners. However, a single best rendition is predicted without reference to additional context.

Without sufficient context, appropriateness is arbitrary. An isolated sentence has multiple valid prosodies with varying frequency of occurrence. Two approaches to determining appropriateness would be: rating appropriateness given a specific context; or collecting contexts that make a given prosody likely (cf. Section~\ref{sec:evaluation-qualitative}~(iii)). However, previous studies  highlight how listener perception of prosody can be affected by both context and what listeners are told to attend to \cite{Cole2014,turnbull2017prominence}. Thus, our current work focuses on generating distinct prosodic renditions and exploring how differences in prosody are perceived, deferring full-scale appropriateness evaluation for future work.

\section{Learning a discrete prosodic representation} \label{sec:discrete-prosody}

We present two methods for learning `\emph{intonation codes}': a baseline using an autoencoder (AE) and k-means in Section~\ref{sec:AE-k-means}; and our proposed method using a variational autoencoder (VAE) with learned multi-modal structure in the latent space in Section~\ref{sec:VAMP}. However, first we address the issue of domain.  

\subsection{Prosodic phrasing}
\label{sec:phrase}

An obvious domain for prosodic control is the prosodic phrase, but accurately locating prosodic phrase boundaries (breaks) requires manual annotation. While there is a correlation between syntactic and prosodic \emph{structure} \cite{syntax-prosody-correlation:2018}, mismatches between syntactic and prosodic phrase boundaries are common \cite{ladd2008ch8}. So, instead we adopt Liberman and Church's notion of \emph{chinks 'n chunks} \cite{chinks-and-chunks:1992} which aims to identify contiguous units of text that map more appropriately to phrases for TTS.\footnote{We thank Oliver Watts for suggesting this method, and helping with the finer details of the parser.}

\emph{Chinks 'n chunks} is a simple heuristic parser that takes advantage of the right-branching nature of English; content words tend to occur towards the end of phrases and function words towards the beginning. However, since certain word types can behave like either, Liberman and Church define two categories:

\vspace{-6pt}
\begin{center}
\begin{tabular}{c@{\quad---\quad}c@{\quad+\quad}c}
    \textbf{chink} & function words & tensed verbs \\
    \textbf{chunk} & content words & objective pronouns
\end{tabular}
\end{center}
\vspace{-6pt}
Tensed verbs can behave like auxiliaries, thus starting a phrase. Objective pronouns can behave like nouns, thus acting as content words. The parsing algorithm is a simple greedy match of \textbf{\{chink* chunk*\}}, see Table~\ref{tab:sentence-stats} for phrase examples (in bold).

\begin{figure}[t]
  \centering
  \includegraphics[width=\linewidth]{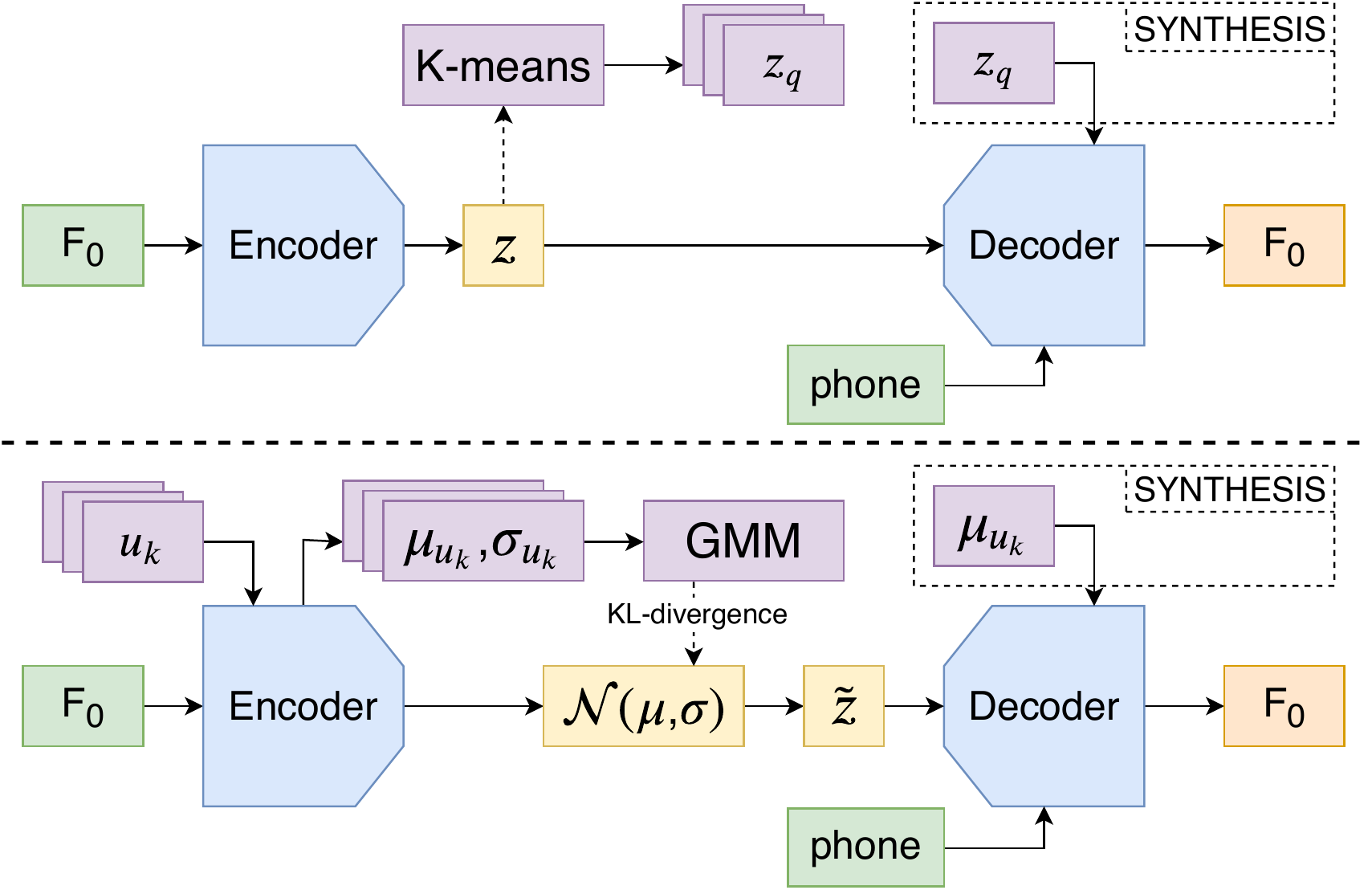}
  \caption{Architecture for \kmeans{} (top) and \vamp{} (bottom). Yellow and purple indicate phrase-level portions, while purple shows specifically where discreteness is added.}
  \label{fig:systems}
  \vspace{-12pt}
\end{figure}

\vspace{-2pt}
\subsection{Baseline: two-stage clustering} \label{sec:AE-k-means}

Our baseline (top of Figure~\ref{fig:systems}) has two stages: learn continuous embeddings $z$ using an AE; cluster the training data embeddings using k-means. We call the clusters $z_q$ `intonation codes'.

Note that an AE's reconstruction loss (indirectly) encourages similar inputs to locate close to each other in the embedding space, thus imposing an implicit distance (without scale). For this reason, unsupervised clustering is feasible, though the two-stage process may lead to sub-optimal \emph{intonation codes}.

\vspace{-4pt}
\subsection{Probabilistic multi-modal latent space} \label{sec:VAMP}

The two-stage approach in Section~\ref{sec:AE-k-means} is limiting as the AE will not necessarily structure its space into clusters. Ideally, the embedding space and clustered structure would be learnt jointly.

VAEs \cite{VAE:2013} are subject to a prior reflecting our assumptions about the underlying latent factors that describe the data. This prior directly enforces distance and scale on the space, which can in turn be used to enforce a clustered structure. In \cite{VAE:2013}, a unimodal Gaussian is used, but we want to find distinct prosodic behaviours in our data. Hence, we use a variational mixture of posteriors (VAMP) prior \cite{VampPrior-VAE:2018} (bottom of Figure~\ref{fig:systems}, in purple).

In simple terms, the VAMP prior is a Gaussian mixture model (GMM), whose parameters are learned jointly with the rest of the model. However, we do not learn the GMM parameters directly, instead we learn $K$ `\emph{pseudo-inputs}' $\{\uu_k\}_{k=1}^K$, where $K$ is a hyperparameter. These \emph{pseudo-inputs} are not real inputs, they are parameters learned through backpropagation. Each GMM component is given by a \emph{pseudo-input}'s approximate posterior $p(\z_k \mid \uu_k) = \mathcal{N}(\z_k; \mub_{\uu_k}, \sigmab_{\uu_k})$. Here, we define our \emph{intonation codes} using GMM component centres $\mub_{\uu_k}$.

Since we learn \emph{pseudo-inputs} and not GMM parameters to define our prior, we are learning parameters in the input space. Tomczak and Welling \cite{VampPrior-VAE:2018} demonstrated this for fixed-size images; we present what we believe to be the first application of VAMP to sequence data (F0 contours). Therefore, we have to contend with learning a sequence of parameters for each \emph{pseudo-input}. While it may be possible to learn the sequence lengths, in this work we choose to fix the number of frames of each \emph{pseudo-input} at initialisation. See Section~\ref{sec:systems} for more discussion on \emph{pseudo-input} sequence length.

\vspace{-4pt}
\section{Data} \label{sec:data}

Our choice of training data is motivated by the need for interesting variation: if the data is very stylistically consistent, there will be too little variation to capture through \emph{intonation codes}. We therefore use the Blizzard Challenge 2018 dataset \cite{blizzard:2018} consisting of stories read in an expressive style for a 4--6 year old audience. In total it contains 6.5 hours (\texttildelow7,250 sentences) of professionally-recorded speech from a female speaker of standard southern British English. Three stories were held out for the listening test: Goldilocks and the Three Bears, The Boy Who Cried Wolf, and The Enormous Turnip.

While this is not conversational data, it does contain character voices and direct speech. Our \emph{intonation codes} may capture the child audiobook style as opposed to prosody typically seen in dialogue. However, this work serves as a proof of concept that we will later validate using dialogue data \cite{eva-dialogue-data:2019,goodhue2016bestiary}.

\section{System details} \label{sec:systems}

Our two models,\footnote{Code is available at \href{https://github.com/ZackHodari/discrete_intonation}{github.com/ZackHodari/discrete\_intonation}} \kmeans{} and \vamp{}, both have an auto-encoder structure, encoding and reconstructing mean-variance normalised logF0, delta, and delta-delta features. MLPG \cite{MLPG:2000} is used for F0 generation using global standard deviation. This F0 contour is then synthesised with natural spectral features using WORLD \cite{WORLD:2016} and a frame-shift of 5ms. For TTS the \emph{intonation codes} for the decoder must be chosen without using natural F0, as discussed in Section~\ref{sec:evaluation}.

The encoders and decoders for both systems are as follows: a feedforward layer with 256 units, followed by three recurrent layers using gated recurrent cells with 64 units. Finally, outputs are projected to the required dimension. Both decoders are conditioned on one-hot phone identity. We found that a full linguistic specification limited the range of variation captured in F0. Phone identity was upsampled to frame-level using forced alignment durations.\footnote{Equivalent to step-wise hard monotonic attention \cite{marginalisation-hard-attention:2019,hard-stepwise-monotonic-attention:2019} in a sequence-to-sequence model. In the future we'll use an encoder with attention to utilise the learned prosodic features of these models \cite{eva-spontaneous:2019}.} The encoders are clocked at the frame-level, so to get the sequence of phrase-level \emph{intonation codes} for a sentence, we take the encoder outputs at the last frame of each phrase, and assign each output to a cluster/mode. The \emph{intonation codes} are defined as follows:

\begin{enumerate}[leftmargin=1.8cm,itemsep=2pt,topsep=5pt,parsep=0pt,partopsep=0pt]
    \item[\kmeans{}] -- $\z_q$ (cluster centroids)
    \item[\vamp{}] -- $\mub_{\uu_k}$ (mean of \emph{pseudo-input} approx.\ posteriors)
\end{enumerate}

We use 20 clusters for \kmeans{}, and 20 \emph{pseudo-inputs} for \vamp{}. As discussed earlier, we fix the sequence length of the \emph{pseudo-inputs} at initialisation. Using the same sequence length for all \emph{pseudo-inputs} was adequate and gave a stable model, if that sequence length is within the range seen in the training set: \texttildelow50 to \texttildelow500 frames. However, we obtained more distinct clusters by using varied \emph{pseudo-input} sequence lengths. We used sequence lengths from 50 to 500 frames, with a step of 50 and repeating each length twice, for a total of 10 unique sequence lengths, and 20 \emph{pseudo-inputs}. We used each sequence length twice to allow for multiple modes at each length.  

Both models were trained for 100 epochs using Adam \cite{adam:2014} with a learning rate increasing linearly from 0 to 0.005 over the first 8 epochs and then decaying proportional to the inverse square of the number of batches \cite[Sec~5.3]{transformer:2017}. Our batch size is 32. The KL-divergence term in \vamp{} is weighted by zero during the first 5 epochs and increased linearly to 0.001 over 20 epochs. \vamp{} converged to a KL-divergence of 5.32. When using the oracle embedding, \kmeans{} and \vamp{} achieved F0 RMSEs of 33.0Hz and 37.1Hz, respectively.

\section{Evaluation} \label{sec:evaluation}

Recall that we aim to capture distinct prosodic characteristics using \emph{intonation codes}, such as changes affecting information structure or strength of expressivity. Therefore the first step in evaluation is to determine whether changing the \emph{intonation code} produces perceivable variation. Generating a new rendition of a sentence requires selecting a sequence of \emph{intonation codes}: one per (\emph{chink 'n chunk}-based) prosodic phrase. While both systems learn using multi-phrase sentences, we do not have a ``language model'' over these \emph{codes}, as such we cannot know which \emph{code} sequences are appropriate. Thus, we restrict the current work to sentences with one phrase and leave multi-phrase synthesis for future work. We randomly chose 12 single-phrase test sentences: 4 from each of the test set books in Table~\ref{tab:sentence-stats}.

\subsection{Subjective evaluation} \label{sec:evaluation-subjective}

\begin{figure}[t]
  \centering
  \includegraphics[width=\linewidth]{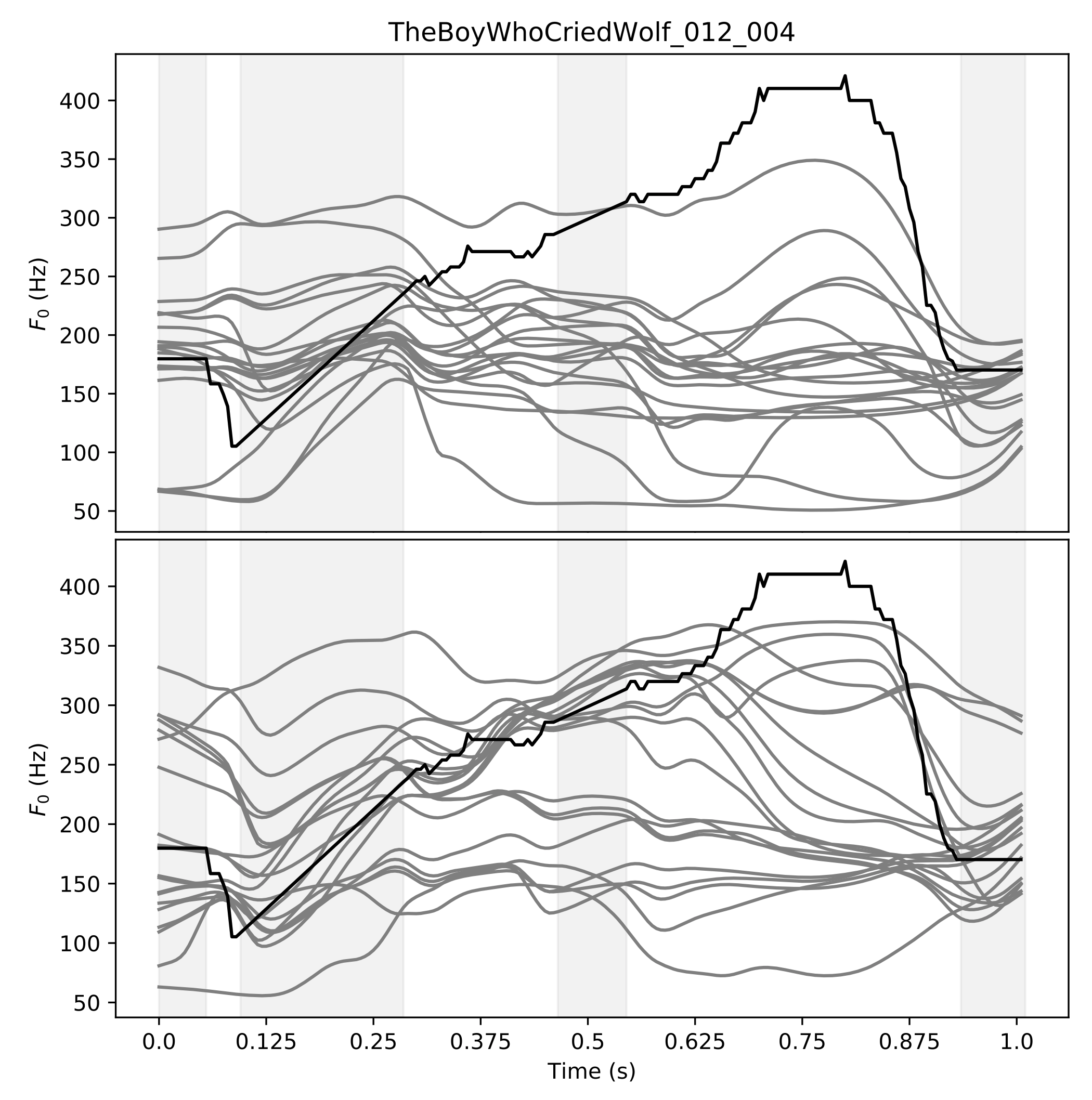}
  \caption{20 \emph{codes} for \kmeans{} (top) and \vamp{} (bottom) for the sentence: ``What's the matter now?''. The black line shows natural F0, interpolated linearly in unvoiced regions.}
  \label{fig:repeated-f0}
  \vspace{-6pt}
\end{figure}

In a forced choice listening test, listeners were presented with two renditions of the same sentence and asked if they had ``different intonation''. We synthesised 40 renditions (20 \kmeans{} clusters + 20 \vamp{} modes; Figure~\ref{fig:repeated-f0}) of each of the 12 test sentences, from which we randomly chose 38 pairs. Each pair comprised two different renditions of the same sentence, both from the same system. A 2x2 Latin Square between-subjects design was used so that each listener heard all sentences, half the pairs from \kmeans{} and half the pairs from \vamp{}. Across two listeners all pairs were presented once. 22 native English-speaking participants each took around 45 minutes to complete the test, for which they were paid \pounds8.

\begin{figure}[t]
  \centering
  \includegraphics[width=\linewidth]{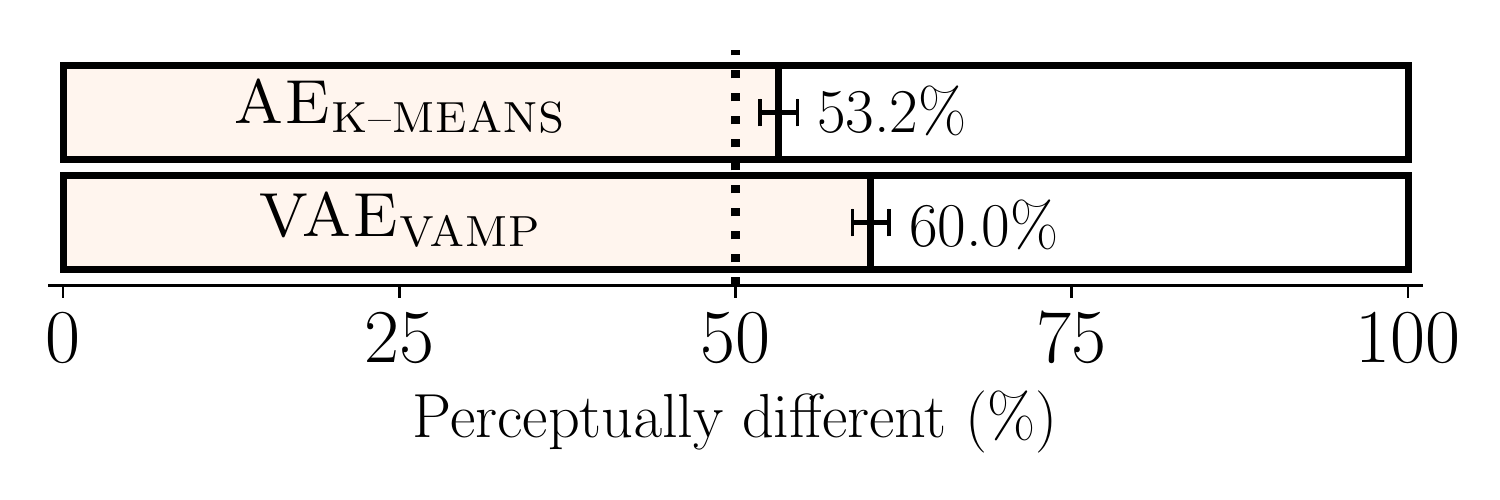}
  \caption{Same/different results. Error bars shows binomial confidence interval.}
  \label{fig:same-different-systems}
  \vspace{-12pt}
\end{figure}

Taking the results per system in a binomial significance test, we find that overall each system produced significant perceptual differences (Figure~\ref{fig:same-different-systems}). The rate of perceptual difference for \vamp{} was significantly more than for \kmeans{}. Taking results per pair, we performed binomial significance tests for the 38 pairs of both \kmeans{} and \vamp{}, followed by Holm-Bonferroni correction over all 76 pairs. After the correction, 10 pairs for \kmeans{} and 16 pairs for \vamp{} showed significant perceptual difference
(corrected $p < 0.005$).

\subsection{Qualitative evaluation} \label{sec:evaluation-qualitative}

While we have shown that \vamp{} produces distinct renditions more frequently, the above evaluation does not reveal \emph{how} they differ, or what the induced \emph{intonation codes} have captured. To understand this, we first need to know whether our distinct renditions are interpreted differently and if so, how.

Ideally, listeners would identify prosodic constructions linked to specific interpretations, and we would then examine their distributions across \emph{intonation codes}.  However, limiting this to previously identified categories or constructions (e.g., \cite{ward2019book, goodhue2016bestiary}) risks missing important types of variation captured by the model. Similarly, a narrow focus on specific linguistic or affective phenomena also increases difficulty for non-expert listeners, and potentially introduces bias. So, to explore this space, we carried out a small, qualitative study. More specifically, we explored: \textbf{(i)} whether the prosodic differences captured discourse/information structural or affective differences in meaning, \textbf{(ii)} whether \emph{intonation codes} were interpreted in a consistent way across sentences, and \textbf{(iii)} what types of variation in prosodic meaning are salient to non-expert listeners.

We took the 6 \vamp{} pairs with the largest percentage of ``different intonation'' judgements in the previous test; the mean across listeners for each of these pairs ranged from 76.5\% to 82.6\%. We ran 45 minute one-on-one interviews with 5 native English-speaking participants (paid \pounds8). We exclude this first (pilot) interview results from the following analysis. We asked listeners to comment on how the sentence was performed and, what effect it had---e.g.\ did the meaning or emotion change? During the interview, listeners were given all 12 sentences for one \emph{code} pair at a time (i.e.\ 2 renditions for each of the 12 sentences), and were able to choose which renditions to comment on. Some chose to compare two renditions of a sentence, while others discussed individual renditions independently.

\vspace{-8pt}
\paragraph*{(i)}
We summarised the interview transcriptions by categorising comments according to descriptive terms; out of 68 total terms, 26 were used to describe more than one sentence. The terms used to describe 4 or more sentences (in order of frequency) were: upset, statement, narrative, question, surprised, ``standard'' style, continuation rise,\footnote{This term was not used directly, but listeners described the effect.} emotional, anticipatory, sad, child storytelling, monotonous, and confused. The broad range of terms used is testament to the variety of prosodies our \emph{intonation codes} have captured. However, most of the terms related to more affect-related changes, which is not so surprising given our data. Changes in interpretation relating to information/discourse structure were reported, most notably continuation rise. However, many other stance/interaction related descriptions were also given, e.g. back-channelling, insincere apology/impressed/surprise; and humorous/typical sarcasm.

\vspace{-8pt}
\paragraph*{(ii)}
Certain \emph{intonation codes} were consistently reported to produce styles such as: questioning, upset, and narrative. In some cases a style was described, but noted as inappropriate (most notably, questioning). Nonetheless, \emph{codes} were not wholly consistent, with their interpretation often changing depending on the sentence. Table~\ref{tab:sentence-stats} shows the number of unique terms, and the terms used multiple times for each sentence. This demonstrates that, unsurprisingly, semantics has a large impact on the perceived effect of the \emph{codes}. The least descriptive sentences, such as ``What's the matter now?'' and ``I'm sorry'', elicited the most comments from listeners. This is either because our \emph{intonation codes} are able to produce more variation more freely, or because listeners can imagine more contexts for them. In order to fully determine if individual \emph{codes} behave consistently, we would need a larger sample, and should design sentences specifically for the test.

\begin{table}[t]
    \begin{center}
	\renewcommand{\arraystretch}{1.1}
    \caption{Single-phrase test sentences: the total number of unique terms used to describe each sentence, and lists of terms used more than once for each sentence.}
    \resizebox{\linewidth}{!}{\begin{tabular}{ r|p{9.25cm} }
        13 & \textbf{There was no answer.} --- statement, upset, surprised, anticipatory \\
        11 & \textbf{``I'm so hungry.''} --- upset, statement, continuation rise \\
        15 & \textbf{``Too hard!''} --- question, statement \\
        10 & \textbf{They climbed the stairs.} --- upset, continuation rise, anticipatory, sad, narrative \\
        \multicolumn{2}{c}{} \vspace{-3mm} \\
        
        20 & \textbf{``What's the matter now?''} --- statement, question, rhetorical, annoyed, friendly, urgent \\
        11 & \textbf{``We'd better make sure.''} --- upset, question, ``standard'' style, uncertain \\
        12 & \textbf{``Do you think we're so stupid?''} --- insulted, upset, rhetorical, sad \\
        19 & \textbf{``I'm sorry.''} --- fake apology, passive aggressive, question, apology, ``standard'' style, upset \\
        \multicolumn{2}{c}{} \vspace{-3mm} \\
        
        9 & \textbf{He wanted a turnip.} --- statement, narrative, continuation rise, sad, bored \\
        7 & \textbf{They both tugged and tugged.} --- narrative, upset, child storytelling, ``standard'' style \\
        11 & \textbf{But the turnip didn't move.} --- upset, statement, narrative, surprised \\
        14 & \textbf{``It's enormous!'' cried Jack.} --- surprised, exclamation, childlike
    \end{tabular}}
    \label{tab:sentence-stats}
    \end{center}
    \vspace{-12pt}
\end{table}

\vspace{-8pt}
\paragraph*{(iii)}
In general, interpretations often appeared dependent on what contexts listeners thought were appropriate for a specific rendition. In fact, some listeners provided rich descriptions of contexts a rendition might make sense in. This could be a useful direction for analysing what a learned representation captures. We could conduct one-on-one interview where listeners are asked to describe some context a rendition might make sense in---selecting ``unsure'' or ``invalid'' when necessary. From this descriptive task we could categorise interpretation of different renditions and determine if renditions consistently correspond to plausible, and potentially uncommon, contexts.

Interestingly, users perceived some duration and loudness changes, though neither of these features were modified.

\section{Conclusion} \label{sec:conclusion}

We presented a discrete prosodic representation that operates in the phrase domain and produces multiple perceptually distinct renditions of individual sentences. We observed a broad range of affective, and some information structural variation. The interpretation of renditions varied based on semantics, where ambiguity lead to users inventing contexts based on what they perceived. This lead to a new idea for better evaluating the perceived effect of different prosodic renditions, using an interview-based descriptive task.

{\footnotesize \noindent \textbf{Acknowledgements:} Zack Hodari was supported by the EPSRC Centre for Doctoral Training in Data Science, funded by the UK Engineering and Physical Sciences Research Council (grant EP/L016427/1) and the University of Edinburgh. We thank Oliver Watts for his support both with prosodic phrasing and advice when training \vamp{}.}

\bibliographystyle{IEEEtran}
\bibliography{references,prosody}

\end{document}